\documentclass[runningheads]{llncs}
\usepackage{graphicx}
\usepackage{multirow}
\usepackage{lipsum}
\usepackage{subcaption}
\usepackage[para]{footmisc}
\usepackage[normalem]{ulem}
\useunder{\uline}{\ul}{}
\usepackage{booktabs}
\usepackage{amsmath}
\usepackage{color}
\usepackage{algorithm}
\usepackage{algpseudocode}
\newcommand{\cm}[1]{\textcolor{black}{#1}}
\newcommand{\maria}[1]{\textcolor{black}{#1}}

\let\ul\relax

\usepackage[noadjust]{marginnote}
\newcommand{\pageenlarge}[1]{\enlargethispage{#1\baselineskip}}

\DeclareMathOperator*{\argmax}{arg\,max}

\usepackage[show]{chato-notes}
\usepackage{adjustbox}

\begin{document}
\title{What Else Would I Like?  A User Simulator using Alternatives for Improved Evaluation of Fashion Conversational Recommendation Systems}
\titlerunning{What Else Would I Like?}

\author{Maria Vlachou \and
Craig Macdonald}
\authorrunning{Maria Vlachou \& Craig Macdonald}
\institute{University of Glasgow, Glasgow, UK \\
\email{m.vlachou.1@research.gla.ac.uk},\\
\email{craig.macdonald@glasgow.ac.uk}}
\maketitle              %
\begin{abstract}
\looseness -1 In Conversational Recommendation Systems (CRS), a user can provide feedback on recommended items at each interaction turn, leading the CRS towards more desirable recommendations. Currently, different types of CRS offer various possibilities for feedback, i.e., natural language feedback, or answering clarifying questions. In most cases, a user simulator is employed for training as well as evaluating the CRS. \cm{Such user simulators typically critique the current retrieved (items) based on knowledge of a single {\em target} item.} Still, evaluating systems in offline settings with simulators suffers from problems, such as focusing entirely on a single target item (not addressing the exploratory nature of a recommender system), and \cm{exhibiting extreme patience (consistent feedback over a large number of turns}). 
To overcome these limitations, we obtain extra judgements for a selection of alternative items in common CRS datasets, namely Shoes and Fashion IQ Dresses. Going further, we propose improved user simulators that allow simulated users not only to express their preferences about alternative items to their original target, but also to change their mind and level of patience. In our experiments using the relative image captioning CRS setting and different CRS models, we find that using the knowledge of alternatives by the simulator can have a considerable impact on the evaluation of existing CRS models, specifically that the existing single-target evaluation underestimates their effectiveness, and when simulated users are allowed to instead consider alternatives, the system can rapidly respond to more quickly satisfy the user.

\end{abstract}
\section{Introduction}
\begin{figure}[tb]
\centering
\includegraphics[width=10cm]{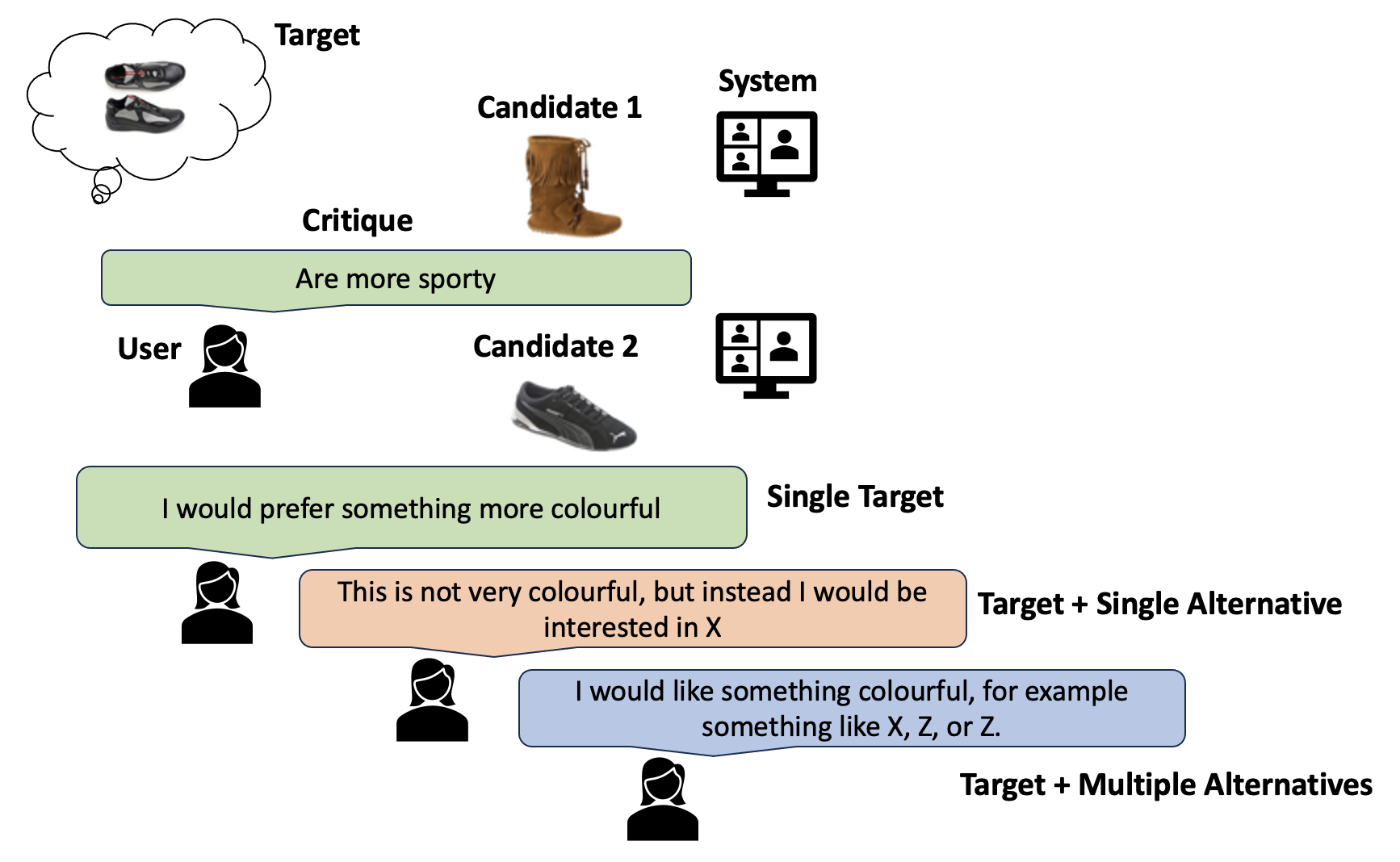}
\caption{Example of a fashion Conversational Image Recommendation scenario. At each turn, the user provides natural language feedback on a candidate item. In existing systems, users are assumed to have a specific target in mind (green). Instead, the presence of a single alternative (orange) or multiple alternative (blue) items can guide the system to find a target of a certain type.}\label{fig:example_alt}\vspace{-\baselineskip}
\end{figure}

In recent years, online shopping is becoming increasingly popular, leading to the development of e-shopping platforms such as Amazon, which help users with product search~\cite{rowley2000product,zou2019learning}. Due to the prevalence of online services and personal digital assistants~\cite{brill2019siri}, research in fashion recommendation is gaining ground. As a result, various approaches and related datasets have been proposed that focus mainly on image recommendation and retrieval~\cite{han2017learning,he2016ups,liu2016deepfashion,mcauley2015image}. The influence of \cm{fashion-related research} in the recommender systems community is also reflected in recent \cm{fashion-related} initiatives such as the {\em Workshop on Recommender Systems in Fashion\footnote{\url{www.fashionxrecsys.github.io/fashionxrecsys-2023/}}} and the {\em RecSys Challenge 2022\footnote{\url{www.recsyschallenge.com/2022/}}}. Due to the highly personalised nature of fashion recommendation, {\em conversational} recommendation systems (CRS) -- which allow users to express their feedback as  natural language -- have gained importance~\cite{kovashka2013attribute,kovashka2017attributes,yu2017fine} \cm{in this domain}. In \cm{such settings}, users' preferences are collected over a number of interaction turns with the system. This is described as {\em interactive image retrieval} in a {\em Conversational Image Recommendation} setting, and the user natural language feedback is a {\em critique} of the relative differences between the user's desired item and the recommended image item at each turn~\cite{guo2018dialog,wu2020fashion,wu2021partially,wu2022multi,yu2019visual}. An example of an user interacting with a CRS is shown in Figure~\ref{fig:example_alt}. Here, the user provides textual feedback regarding the suggested item of the previous turn (green), by focusing on a defined target. Currently, there is no option to request for one (orange) or more (blue) alternatives. Note that each {\em candidate} item shown to the user at each turn is simply the top-ranked item among a ranking of items. The training and evaluation of CRSs is often based on the interactions of a {\em simulated} user, which has a desired target item ``in mind". This is suggestive of a known-item type of task~\cite{broder2002taxonomy}, where the target item is assumed to exist in the item catalogue. 

However, this might not always be true. Indeed, the current fashion recommendation evaluation setting presents some limitations: First, the {\em realisticity} of an interaction does not necessarily aid user experience. Specifically, the user is assumed to be infinitely patient, and willing to interact with the CRS for a large number of turns until the target item is found. This setting is not representative of a real user experience, where a user might become frustrated. To complement this, a simulated user is assumed to be single-minded, meaning that it is not flexible enough to change its strategy or initial plan. On the contrary, recommender systems are typically used to aid exploratory user behaviour~\cite{broder2002taxonomy,o2006race}, and therefore, by persisting on a single desired item, users are not exploring the product space. Moreover, unlike information retrieval systems, which are evaluated using test collections that aim to provide a reasonably {\em complete} coverage of relevant documents, recommender systems suffer from a lack of completeness. In particular, search engines use {\em pooling} of documents retrieved from a various systems and a per query {\em relevance judging} of pooled documents to obtain \cm{more complete assessments}. For example, the MSMARCO test collection has thousands of queries contains shallow judgements, while the TREC Deep Learning track provided $\sim$100 queries with deeper judgements~\cite{craswell2020overview}. In this regard, it was found that the sparse MSMARCO assessments are not a suitable replacement for more complete assessments~\cite{macavaney2023one}. Similarly, the presence of more reliable relevance judgments for CRS would benefit the reliability of their evaluation. In what follows, we aim to show that more relevance judgments for CRS items can be obtained by directly asking users about their alternative preferences, thus allowing the user to update their preferences during the dialogue.

In short, this work contributes: (i) Extended datasets for two fashion-related CRS datasets that contains labels about the presence of sufficient alternatives for a number of known target items by real users on different fashion item categories. In this way, we contribute to evaluation completeness for relevance; (ii) A meta-user simulator, which wraps an existing user simulator to provide feedback for possible alternatives items; and (iii) a study of existing CRS models evaluated with and without alternatives. We find that the existing single-target evaluation of CRS underestimates their effectiveness, and when simulated users are allowed to instead consider alternatives, the system can rapidly respond to more quickly satisfy the user.
The rest of the paper is organised as follows: We present some related work on Conversational Recommendation, simulators, and data pooling in Section~\ref{sec:rw}, we introduce our alternatives-based user simulator in Section~\ref{sec:method}, and our alternatives extended datasets in Section~\ref{sec:dataset}. Experiments and conclusions follow in Sections \ref{sec:results} \&~\ref{sec:conc}.

\section{Related Work}\label{sec:rw}
\subsection{Conversational Recommendation Systems (CRS)}\label{ssec:crs_rw}
The main goal of Conversational Information Systems~\cite{zamani2023conversational} is to assist users with completing tasks~\cite{chen2017survey,serban2018survey}, such as finding directions or buying products using dialogue systems~\cite{sun2018conversational,trippa2019spoken}. For example, in Conversational Search~\cite{aliannejadi2019asking,azzopardi2018conceptualizing,bi2019conversational,radlinski2017theoretical,zhang2018towards}, users seek information and respond to the system's clarifying questions. However, our interest lies in Conversational Recommendation Systems (CRS)~\cite{christakopoulou2016towards,Jannach_2021,lei2020estimation,liu2020towards}, which help users find items of interest by providing feedback on suggested items and responding to clarifying questions regarding their preferences. In recent years, many CRS models use a {\em reinforcement learning (RL)} policy network, where at each turn, an action is taken based on transition probabilities between states~\cite{zhang2020evaluating,guo2018dialog,vakulenko2019qrfa,wu2021fashion}. Indeed, RL optimizes for long-term rewards~\cite{shi2019build}, i.e.\ retrieving the correct item in later turns, and is therefore, assumed to improve long-term system performance. In contrast, other approaches focus on generating clarifying questions about candidate items or attributes until the system gets a clear picture of users' preferences, before making a final a recommendation~\cite{christakopoulou2016towards,zhang2018towards,zou2020towards}. Finally, some systems combine different characteristics, for example by separately modeling feedback during a dialogue and historic feedback and choosing between recommendations and clarifying questions at each turn~\cite{ren2021learning,ren2022variational}.

In contrast to purely text-based recommendation, a further type of CRS \cm{is} {\em Conversational Image Recommendation}~\cite{guo2018dialog} in the fashion domain. The main differences from text-based CRS are that: (a) recommended items are displayed as images, and more specifically, a user only sees the top-ranked image item at each turn, and (b) the setting does not differentiate between user feedback and providing the user need, since the user feedback is provided in natural language form and describes specific attributes of the desired item. In particular, the feedback \cm{sought from the user is sought to describe} the relative visual differences between the candidate and the target image, which aims to lead to an improved recommendation at the next turn. As shown in Figure~\ref{fig:example_alt}, a user \cm{is assumed to have} a  target item in mind, and provides natural language feedback at every turn. Such systems use a gated recurrent unit (GRU) model~\cite{guo2018dialog,hidasi2015session,yu2020towards} or a gated recurrent network (GRN) to separately process the textual feedback and visual recommendations for multi-modal sequence representation~\cite{wu2022multi}, while others place more importance on long-term user satisfaction and historical feedback~\cite{wu2021partially}. Additionally, \cm{a variant of this task may allow the user to} have the opportunity to provide feedback on a list of candidate items~\cite{yu2019visual}.

The common assumption in these approaches is that the dialog with the user proceeds with a  narrowly-defined target item. However, by being single-minded and not allowing for any other option, the user \cm{may not be adequately able to} explore the product space. Moreover, while displaying the top-ranked images resembles the context of online shopping, a more natural conversation usually involves a user that changes their mind, and does not wish to interact for an infinite number of turns. \maria{In contrast}, in our work, we allow the user to reconsider after a certain threshold and provide alternative options.

\subsection{User Simulation for Evaluating CRS}\label{ssec:sim}
Training CRS systems in a multi-turn setting requires a large amount of data~\cite{li2016user,shi2019build}. To compensate for the increased need for real users, user simulators are used as a surrogate of human behaviour~\cite{li2016user,shi2019build}. Indeed, several approaches have been proposed that employ user simulators in interactive systems~\cite{chung2004developing,griol2013automatic,owoicho2023exploiting,sun2023metaphorical,verberne2015user,zhang2020evaluating,zhang2022analyzing}. For example, Owoicho et al.~\cite{owoicho2023exploiting} observed improved performance of mixed-initiative conversational search systems with multiple rounds of simulated user feedback, while Sun et al.~\cite{sun2021simulating} simulated user satisfaction with training data \cm{from} annotators who judged the level of satisfaction of each turn from the dialogue context. As for CRS, recent work on user simulators builds on an agenda-based framework that uses push and pull operations to update the user needs per turn~\cite{balog2021conversational,schatzmann2007agenda,vakulenko2019qrfa,zhang2020evaluating}. For Conversational Image Recommendation, the state-of-the-art simulator framework is {\em relative captioning}, where the dialog system is trained on human-annotated captions that describe the visual differences between two image items, and therefore, the feedback can be modified based on the current displayed candidate. 

For evaluating CRS systems, some approaches compare the resulting dialogue with human dialogues using different performance metrics~\cite{sun2021simulating,zhang2020evaluating,wu2021fashion,zhang2022analyzing}. In addition, some simulation approaches collect annotated datasets for training CRS. However, they are usually limited to rating the level of dialogue success or user satisfaction~\cite{sun2021simulating}. On the other hand, our work is focused on extending the completeness of the ground truth by introducing more options to the target space. In other words, we aim to enrich our simulated users with a target {\em group} instead of single target items in a relative captioning setting. In that sense, our approach is similar to Sun et al.~\cite{sun2023metaphorical}, \maria{where they assume an analogical thinking of users, i.e., users comparing new items with prior knowledge. Still, they do not necessarily provide other preference options. Instead, the basis of our work is to inform the user simulator with alternatives.}

\subsection{Data Pooling and Evaluation Completeness}\label{ssec:completeness}
In general, the evaluation of recommender systems is plagued by a lack of completeness, as typically past interactions are ``replayed'' and the prediction ability of the recommender system to predict the hidden ``future'' interaction(s) is measured by classical evaluation measures such as MRR and NDCG. This tends to favour systems that behave similarly to the system originally deployed when the user interactions were collected~\cite{10.1145/3240323.3240370,10.1145/3397271.3401230,10.1145/3458509}. In contrast, search engine evaluation uses test collections~\cite{sanderson2010test}, which combine two techniques for obtaining a more {\em complete} coverage of relevant documents: the {\em pooling} of document retrieved by a number of effective systems; and the explicit judging of the relevance of all pooled documents to a user's query. Incomplete test collections is well known to result in unreliable evaluation~\cite{buckley2007bias,buckley2004retrieval}. Recently, Craswell et al.~\cite{craswell2020overview} found a good correlation between evaluation using thousands of single known relevant queries versus using deeply judged TREC queries, however, pseudo-relevance feedback techniques have been shown to work on the latter but not the former~\cite{wang2023colbert}.

Pooling and assessing is typically not used for recommendation, as the user's exact information needs are not clear. However, for fashion-based CRS, where the user has a target item in mind, we argue that it is possible to ask a 3rd party assessor to consider what other items they may have considered. In this way, we develop more complete test collections for fashion-based CRS (using alternative target items), and a more realistic user simulator that can make use of these alternatives during evaluation.

\section{Simulated Users with Alternatives}\label{sec:method}
In this section, we describe our meta-user simulator for using alternatives: its conception and functionality, and specifically how it uses the knowledge of alternatives to inform the CRS about user preferences. In addition, we provide some explanations about the actions of the simulator and the system. 

We firstly start by providing some general notation providing the general principles of a user simulator. A User Simulator using relative captioning is an object with a single function, which takes as input the user's id, the current top-ranked image, $top\_ranked$, and the user's actual target image $target$. It then calls a learned relative captioning model, which has been trained given $target$ to critique $top\_ranked$, and return a text string describing the visual differences between  $top\_ranked$ and $target$:
$$
Usersim.critique(user, top\_ranked, target) \rightarrow \mathcal{T}
$$

Our intuitions for a user that considers alternative items are that (I1) their patience critiquing for a single target item may run out after number of turns, (I2) that when selecting an alternative as a new target, they are influenced by the current item they see, and (I3) the existing relative captioner-based user simulator can be called with a new target.

\looseness -1 Pseudocode for our meta-user simulator procedure is provided in Algorithm~1. More specifically, our $MetaUserSim$ firstly requires knowledge of all possible alternative items for target items. This is akin to the ``qrels'' in test-collection based evaluation. Then, when the meta-user simulator is called, if the turn number exceeds the patience $tolerance$ parameter, then alternatives are considered (line 2, addressing in intuition I1); Among all of the alternatives for a given target, we select the alternative that is closest in image similarity to the current top ranked image as the target (lines 5 \& 6, addressing I2). The existing relative-captioning based user simulator is then asked to critique the newly selected target (line 8, I3); Note that we choose to consider the target as part of the alternatives, so that the ranker can make a choice between the nearest item at a later stage. Finally, we instrument Algorithm~1 to provide data about how often alternative are chosen.

\begin{algorithm}
\caption{Meta-User Simulator}
\label{alg:next_k}
\begin{algorithmic}[1]
\Require{$Usersim$: base user simulator for Conversational Image Recommendations}
\Require{$sims$: function to find the similarity of a set of image from a given query image}
\Require{$tolerance$: patience parameter referring to the turn a user starts to ask for an alternative}
\Require{$all\_alternatives$: A set of known alternative relevant images for all targets}
\Procedure{MetaUserSim.critique}{$turn, top\_ranked$, $target$}

        \If{$turn > tolerance$}
            \State $alternatives = all\_alternatives[target]$
            \State $alternatives.\text{target}(target)$
                \State $all\_dists = sims(alternatives, top\_ranked)$
            \State $target = alternatives[\argmax(all\_dists)]$
        \EndIf
        
        \State \textbf{return} Usersim.critique($turn, top\_ranked, target$)
\EndProcedure
\end{algorithmic}
\end{algorithm}

In the next section, we discuss how our dataset with alternatives is created.

\section{Enriching of CRS Datasets with Alternatives}\label{sec:dataset}

We now describe how we enriched two fashion CRS datasets with alternative judgements. 

\subsection{Original Datasets}\label{ssec:data}
We use the Shoes~\cite{berg2010automatic,guo2018dialog} dataset, which contains 4658 test target images and one relative critique per candidate-target pair, and the FashionIQ Dresses dataset~\cite{wu2021fashion}, which contains 2454 test images and two relative critiques per candidate-target pair, which describe the relative visual differences between the candidate and target image pairs. We obtain labels of relevance (whether a set of candidate images are a sufficient alternative to a given target image).

\subsection{Target Pooling and Data Collection}\label{ssec:pooling}
We use Amazon Mechanical Turk\footnote{https://www.mturk.com} to gain assessments on alternative items. Participants are instructed that each presented target image item is an item they want to buy. Simultaneously, participants see a set of candidate image items that could be a sufficient alternative to the target which, as instructed, is not available in the catalogue. The task is to select, out of the displayed set of candidates, the ones (if any) that best satisfy the user need as an alternative \cm{to the target item}. Finally, participants are asked to indicate the reason for their selection. Participants are selected based on their location (US, to ensure an adequate level of English) and to the extent they could identify with a person who wears dresses or women's shoes, and are paid based on the rules of Mechanical Turk.

We select the target images by sampling 200 target items from each dataset with varying levels of difficulty (we checked this by conducting a preliminary Query Performance Prediction~\cite{carmel2010estimating,cronen2002predicting} analysis of the sampled items using score-based predictors~\cite{roitman2017enhanced,shtok2012predicting}). We estimate the required number of sampled target images with a power analysis using the correlation values of a recent Query Performance Prediction analysis~\cite{vlachou2022performance}, a significance level (alpha value) of $\alpha = 0.05$, to achieve a power of 90\%. The power analysis estimation provided a number lower than 200 targets for each dataset, but we opted for a sufficient amount of targets.

\looseness -1 \cm{For assessment, we derive a pool of} candidate images for each target by using GRU~\cite{guo2018dialog,hidasi2015session} and EGE~\cite{wu2021partially} (detailed further in Section~\ref{ssec:systems} below). Specifically, we select both the nearest neighbours (in their corresponding image embedding spaces) to the target (60\%) and  their top-retrieved images of the final evaluation turn (40\%).
We place more importance on the nearest neighbours, because we are more interested in the similarity of the images rather than how each CRS model ranks them. Specifically, we use the top-4 ranked nearest neighbours of each target item from each CRS model, and the top-3 top-ranked results from each CRS model at turn 10. This results in a total of 14 candidate images per target item. To ensure no duplicates, we checked how many items overlap between the two CRS models (both for nearest neighbours and retrieved results), and in case of common entries, we replaced them with additional items from lower ranks. \maria{We obtained institutional ethical approval for the study, and we paid participants \$0.63 for each MTurk task (or {\em HIT}) for a total duration of 3 minutes (this above the living wage in our country), making a total cost of the study was \$305 (we rejected only 3 HITs for spammy behaviour)}. For both resulting alternative labelled datasets, there were on average 3.5 identified alternatives per target image. We performed a second round of assessment on 40 target items (10\% from each dataset), and measured assessor agreement. We observed a Cohen's $\kappa$ between the two sets of judgements of 0.87, demonstrating a high level of agreement. In the following, we now analyse three representative fashion CRS systems using the alternatives-based user simulator from Section~\ref{sec:method}, and using the alternatives dataset for 200 target items.

\section{Experiments}\label{sec:results}

In this, we experiment to address the following research questions:

{\bf RQ1} What is the impact of using an alternative-based user simulator on the evaluation of existing CRS models? 
 
{\bf RQ2} What is the impact of patience of an alternative-based user simulator?

{\bf RQ3} Does introducing patience change conclusions about what are the most effective models?

{\bf RQ4} How often do users prefer an alternative over their initial target item?

\subsection{Setup: Conversational Recommendation Systems (CRS)}\label{ssec:systems}
We deploy three existing CRS models using both the original relative captioning single-target evaluation setting and our own meta-simulator with alternatives. 
\begin{itemize}
  \item A GRU model~\cite{guo2018dialog,hidasi2015session} with reinforcement learning, which retrieves 100 top items per turn, and is optimised for maximising short-term rewards.
  \item A GRU variant trained with supervised learning, i.e.\ lacking short-term rewards during training.
  \item EGE~\cite{wu2021partially}, which has the advantage of using the user's historical feedback and the system's historical recommendations.
\end{itemize}
Note that we retain the original training for these models using a user simulator that considered a single target item.

\subsection{Setup: Evaluation Measures}\label{ssec:measures}
Following existing work in CRS, we use classical IR evaluation measures to evaluate the ability of each CRS system to retrieve relevant items. In particular, we measure the ability of the CRS to show the user's desired target at rank 1 (Success Rate @ 1) at each turn of the conversation; Moreover, as there is a ranking of images created at each turn, we use nDCG@10 and MRR@10 to evaluate the presence of target items in the ranking. Following \cite{guo2018dialog,wu2021partially}, we terminate conversations at turn 10; if a target item has been found before turn 1, then all evaluation measures are considered to be equal to 1. Finally, and differing from previous work, we consider the alternatives as relevant items for the purposes of evaluation - in this case, a conversation is successful if any alternative (or the original target) is retrieved (this holds for evaluating the meta-simulator; we still use the classical evaluation setup for the base simulator.

\subsection{RQ1 - Impact of alternative-based user simulator on the evaluation of existing CRS models}
Tables 1 and 2 show the performance of the three CRS models at turns 3, 5 and 10 of a conversation after applying our proposed alternative-based simulator on the Shoes and FashionIQ Dresses datasets, respectively. To indicate the difference in performance, we include the percentage of improvement compared to the traditional setting with the non-alternative user simulator.  We use tolerance at turn 2, which resembles a real shopping scenario.

Overall, we observe improved performance on all three evaluation metrics and both Shoes and Dresses. More specifically, as shown in Table 1, for the Shoes dataset, there are considerable improvements on MRR@10 and Success Rate across all three CRS models. For instance, the highest improvement in both MRR@10 and Success Rate is observed for the GRU-SL model, followed by EGE with only small differences. Surprisingly, for NDCG@10,  we observe negative values, which means that CRS performance does not improveor even drops for GRU-RL when introducing alternatives. If we compare these numbers with the first row of Table 3, the improvements are proportionally related to the initial ranking of systems in the traditional relative captioning setting; the system that is initially performing worst (GRU-SL) is most increased, and the opposite holds for GRU-RL, which improves the least. Improvements are in general greater for MRR@10 and Success Rate than for NDCG@10.

\looseness -1 As for the Dresses dataset, we observe improvements across all three evaluation metrics, especially for the initially worst-ranked system (GRU-SL, see also Table 3). Unlike Shoes, for Dresses we observe a positive difference in all cases, and specifically for MRR@10 and Success Rate, effectiveness is doubled when adding the alternative options. Finally, performance is already improved at turns 3 and 5, which means that when a user switches their behaviour at turn 2, they don't have to wait very long to see an alternative product. To answer RQ1, the impact of using an alternative-based simulator is marked positive when evaluating existing CRS models. This suggests that the previous single-target based user simulators were under estimating the effectiveness of the CRS for real users.

\begin{table}[tb]
\caption{Performance Results of the three CRS models of the Shoes dataset at various turns after applying our meta-simulator. (w/o) Indicates before and (w/) after introducing alternatives. The numbers in brackets indicate the percentage of improvement compared to traditional non-alternative user simulators.}
\resizebox{\textwidth}{!}{
\begin{tabular}{l|ccccccccc}
           & \multicolumn{3}{c}{NDCG@10} & \multicolumn{3}{c}{MRR@10} & \multicolumn{3}{c}{SR@1}       \\ \hline
\multirow{2}{*}{CRS Model}     & \multicolumn{9}{c}{turn}                                                                                               \\ \cline{2-10} 
      & 3           & 5            & \multicolumn{1}{c|}{10}          & 3           & 5           & \multicolumn{1}{c|}{10}          & 3       & 5        & 10      \\ \hline
GRU-SL (w/o)     & 0.178       & 0.201        & \multicolumn{1}{c|}{0.209}       & 0.161       & 0.181       & \multicolumn{1}{c|}{0.196}       & 0.100   & 0.110    & 0.150   \\
GRU-SL (w/)    & 0.205       & 0.252        & \multicolumn{1}{c|}{0.237}       & 0.346       & 0.437       & \multicolumn{1}{c|}{0.495}       & 0.257   & 0.352    & 0.436   \\
\% Improv. & (14.11)     & (22.82)      & \multicolumn{1}{c|}{(12.88)}     & (72.76)     & (82.69)     & \multicolumn{1}{c|}{(86.47)}     & (87.95) & (104.76) & (97.61) \\
\hline
GRU-RL(w/o)     & 0.234       & 0.275        & \multicolumn{1}{c|}{0.303}        & 0.218       & 0.255       & \multicolumn{1}{c|}{0.291}       & 0.150   & 0.180    & 0.240   \\
GRU-RL (w/)    & 0.227       & 0.248        & \multicolumn{1}{c|}{0.230}        & 0.356       & 0.459       & \multicolumn{1}{c|}{0.543}       & 0.257   & 0.368    & 0.489   \\
\% Improv. & (-3.03)     & (-10.32)     & \multicolumn{1}{c|}{(-29.31)}    & (48.08)     & (57.14)     & \multicolumn{1}{c|}{(60.43)}     & (52.58) & (68.61)  & (68.31) \\
\hline
EGE (w/o)      & 0.197        & 0.242        & \multicolumn{1}{c|}{0.277}       & 0.171        & 0.216       & \multicolumn{1}{c|}{0.263}       & 0.080   & 0.140    & 0.210   \\
EGE (w/)      & 0.240        & 0.277        & \multicolumn{1}{c|}{0.286}       & 0.350        & 0.474       & \multicolumn{1}{c|}{0.611}       & 0.236   & 0.384    & 0.552   \\
\% Improv. & (19.68)     & (13.48)      & \multicolumn{1}{c|}{(3.19)}      & (68.71)     & (74.78)     & \multicolumn{1}{c|}{(79.63)}     & (98.73) & (93.13)  & (89.76) \\ \hline
\end{tabular}}
\end{table}

\begin{table}[tb]
\caption{Performance Results of the three CRS models of the Dresses dataset at various turns after applying our meta-simulator. Notation as per Table 1.}
\resizebox{\textwidth}{!}{
\begin{tabular}{l|ccccccccc}
           & \multicolumn{3}{c}{NDCG@10} & \multicolumn{3}{c}{MRR@10} & \multicolumn{3}{c}{SR@1}         \\ \hline
\multirow{2}{*}{CRS Model}       & \multicolumn{9}{c}{turn}                                                                                                 \\ \cline{2-10} 
      & 3            & 5           & \multicolumn{1}{c|}{10}          & 3           & 5           & \multicolumn{1}{c|}{10}          & 3        & 5        & 10       \\ \hline
GRU-SL (w/o)     & 0.071        & 0.078       & \multicolumn{1}{c|}{0.072}       & 0.058       & 0.069       & \multicolumn{1}{c|}{0.068}       & 0.071    & 0.078    & 0.072    \\
GRU-SL (w/)     & 0.131        & 0.139       & \multicolumn{1}{c|}{0.125}       & 0.235       & 0.306       & \multicolumn{1}{c|}{0.353}       & 0.127    & 0.238    & 0.316    \\
\% Improv. & (59.64)      & (55.44)     & \multicolumn{1}{c|}{(53.87)}     & (120.24)    & (126.33)    & \multicolumn{1}{c|}{(135.12)}    & (56.81)  & (100.68) & (125.57) \\
\hline
GRU-RL (w/o)     & 0.073        & 0.088       & \multicolumn{1}{c|}{0.075}       & 0.066       & 0.080       & \multicolumn{1}{c|}{0.074}       & 0.035    & 0.045    & 0.045    \\
GRU-RL (w/)     & 0.110        & 0.121       & \multicolumn{1}{c|}{0.099}       & 0.209       & 0.257       & \multicolumn{1}{c|}{0.269}       & 0.127    & 0.177    & 0.216    \\
\% Improv. & (40.84)      & (31.76)     & \multicolumn{1}{c|}{(27.19)}     & (103.53)    & (105.11)    & \multicolumn{1}{c|}{(113.12)}    & (113.99) & (119.20) & (131.21) \\
\hline
EGE (w/o)        & 0.060        & 0.074       & \multicolumn{1}{c|}{0.085}       & 0.055       & 0.072       & \multicolumn{1}{c|}{0.084}       & 0.060    & 0.074    & 0.085    \\
EGE (w/)        & 0.157        & 0.200       & \multicolumn{1}{c|}{0.225}       & 0.317       & 0.419       & \multicolumn{1}{c|}{0.541}       & 0.233    & 0.327    & 0.472    \\
\% Improv. & (88.48)      & (91.79)     & \multicolumn{1}{c|}{(90.22)}     & (140.24)    & (140.85)    & \multicolumn{1}{c|}{(146.22)}    & (117.35) & (126.18) & (138.70) \\ \hline
\end{tabular}}
\end{table}

\subsection{RQ2 - Impact of patience on alternative-based simulator}
\begin{figure}[tb]
    \centering
     \begin{subfigure}[b]{0.25\textwidth}
     \includegraphics[width=\textwidth]{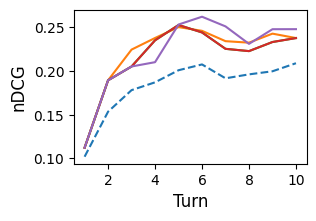}
     \caption{GRU-SL}
     \end{subfigure}
     \begin{subfigure}[b]{0.25\textwidth}
     \includegraphics[width=\textwidth]{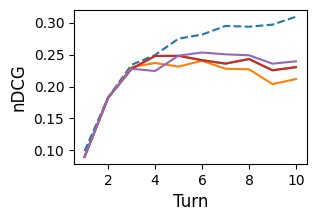}
     \caption{GRU-RL}
     \end{subfigure}
     \begin{subfigure}[b]{0.35\textwidth}
     \includegraphics[width=\textwidth]{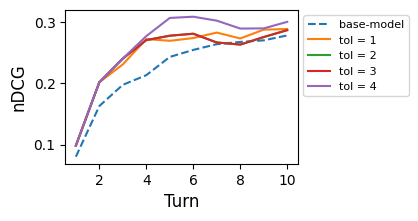}
     \caption{EGE}
     \end{subfigure}
    \caption{nDCG@10 for the various tolerance levels (denoted tol.) before selecting an alternative for the Shoes dataset.}
    \label{fig:shoes_tol_ndcg}
\end{figure}

\begin{figure}[tb]
    \centering
     \begin{subfigure}[b]{0.25\textwidth}
     \includegraphics[width=\textwidth]{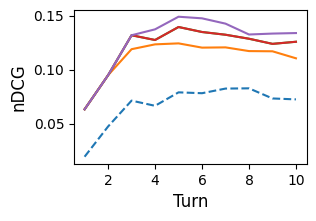}
     \caption{GRU-SL}
     \end{subfigure}
     \begin{subfigure}[b]{0.25\textwidth}
     \includegraphics[width=\textwidth]{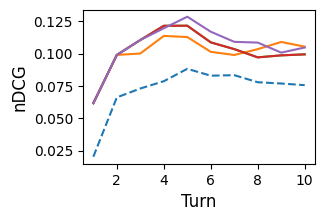}
     \caption{GRU-RL}
     \end{subfigure}
     \begin{subfigure}[b]{0.35\textwidth}
     \includegraphics[width=\textwidth]{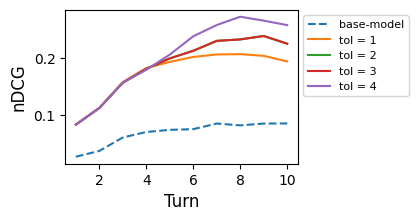}
     \caption{EGE}
     \end{subfigure}
    \caption{nDCG@10 for the various tolerance levels (denoted tol.) before selecting an alternative for the Dresses dataset.}
    \label{fig:dresses_tol_ndcg}
\end{figure}

Figures~\ref{fig:shoes_tol_ndcg} and~\ref{fig:dresses_tol_ndcg} show the NDCG@10 performance at various tolerance levels for Shoes and Dresses, respectively. The solid lines correspond to the different tolerance levels, while the dashed line denotes the baseline evaluation setting without alternatives for each system. In general, the earlier a simulator ``loses" its patience, the earlier the turn there is a boost in performance. However, when tolerance increases (patience lost at later turns), there is a higher performance improvement (compared to the non-alternative simulator) in the long-term. Strangely, for Shoes, we observe a decrease in performance for all tolerance levels for GRU-SL, but in general, values between tolerance levels do not differ significantly. Therefore, the impact of patience is more direct in the turns that follow the tolerance turn, but it is not necessarily different across different levels, thus answering RQ2.

\subsection{RQ3 - Role of patience in the effectiveness of CRS models}
\looseness -1 Table 3 shows the NDCG@10 ranking of the three CRS models  before (first row) and after (remaining rows, each at another tolerance level) introducing the meta-simulator. For both datasets, the ranking changes at tolerance 1 compared with the non-alternative setting, which then remains stable with the varying tolerance levels, with one main difference; for Shoes, there is a re-ranking between the first and the second systems (EGE is ranked first when alternatives are introduced, and GRU-RL moves to the second place), while for Dresses, patience reorders the second and third systems (GRU-SL is improved compared to GRU-RL). To answer RQ3, introducing patience partially changes conclusions about the effectiveness of models, but this change is not further influenced by the level of patience.

\begin{table}[t]
\caption{Resulting ranking (based on NDCG@10) of the 3 CRS models at turn 10 (end of dialogue evaluation setting) using the non-alternative simulator and the various tolerance levels of the alternative-based simulator.}
\begin{adjustbox}{width=\textwidth}
\begin{tabular}{c|c|c}
\hline
Simulator type  & Shoes                                                               & Dresses                                                             \\ \hline
no alternatives & \textbf{GRU-RL}(0.309) $>$ \textbf{EGE} (0.277) $>$ \textbf{GRU-SL} (0.209)  & \textbf{EGE} (0.085) $>$ \textbf{GRU-RL} (0.075) $>$ \textbf{GRU-SL} (0.072) \\
tolerance 1     & \textbf{EGE} (0.288) $>$ \textbf{GRU-SL} (0.237) $>$ \textbf{GRU-RL} (0.211) & \textbf{EGE} (0.194) $>$ \textbf{GRU-SL} (0.110) $>$ \textbf{GRU-RL} (0.105) \\
tolerance 2     & \textbf{EGE} (0.286) $>$ \textbf{GRU-SL} (0.237) $>$ \textbf{GRU-RL} (0.230) & \textbf{EGE} (0.225) $>$ \textbf{GRU-SL} (0.125) $>$ \textbf{GRU-RL} (0.099) \\
tolerance 3     & \textbf{EGE} (0.286) $>$ \textbf{GRU-SL} (0.237) $>$ \textbf{GRU-RL} (0.230) & \textbf{EGE} (0.225) $>$ \textbf{GRU-SL} (0.125) $>$ \textbf{GRU-RL} (0.099) \\
tolerance 4     & \textbf{EGE} (0.300) $>$ \textbf{GRU-SL} (0.487) $>$ \textbf{GRU-RL} (0.239) & \textbf{EGE} (0.258) $>$ \textbf{GRU-SL} (0.133) $>$ \textbf{GRU-RL} (0.104) \\ \hline
\end{tabular}
\end{adjustbox}
\end{table}

\pageenlarge{1}\subsection{RQ4 - Frequency of selecting an alternative}
\begin{figure}[tb]
    \centering
     \begin{subfigure}[b]{0.362\textwidth}
     \includegraphics[width=\textwidth]{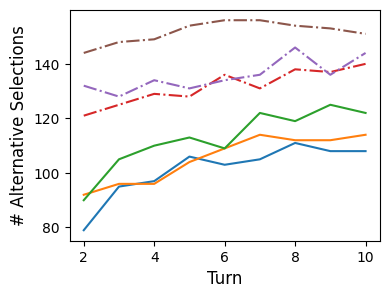}
     \caption{tolerance = 1}
     \end{subfigure}
     \begin{subfigure}[b]{0.53\textwidth}
     \includegraphics[width=\textwidth]{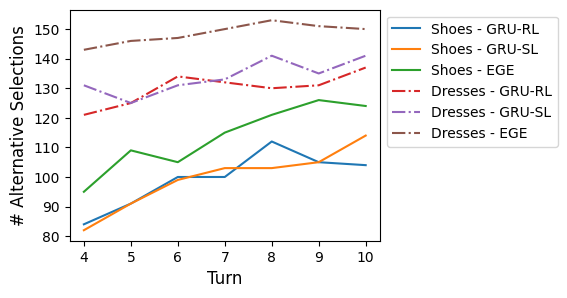}
     \caption{tolerance = 3}
     \end{subfigure}
    \caption{\looseness -1 Number of target images for which the simulator selects an alternative over the target for the three CRS models for tolerance 1 and 3.}
    \label{fig:switch}
\end{figure}

Figure~\ref{fig:switch} shows the number of times, out of the 200 sampled target items, the simulated user opts for an alternative over the initial target, for an earlier (turn 1) and a later (turn 3) tolerance level. The solid lines represent the selection of Shoes CRS models, while the dashed lines denote the performance of Dresses. While the pattern is similar between the two datasets, simulators trained on Dresses tend to select on average more alternative items than simulated users trained on Shoes. This might be explained by the fact that in the initial stage before introducing alternatives, all three systems performed worse when trained on Dresses. Indeed, they might benefit more when there is an option to make an alternative selection. In addition, there is an increased tendency to select alternative items as turns increase and the user patience decreases, which seems reasonable. Finally, we see that the selection of alternatives is more frequent for the EGE model for both datasets compared to the two GRU-based models. In general, observing the high frequency of alternatives indicates that the lower performance of CRS models cannot be solely attributed to their retrieval ability, but also on the lack of sufficient target items. In short, users often tend to pick an alternative when they have the option to do so, thus answering RQ4.

\pageenlarge{2}\section{Conclusions}\label{sec:conc}
\looseness -1 We have introduced a meta-user simulator for improved evaluation of Fashion CRS with alternative relevant items. Our simulator informs the existing simulator with knowledge of alternative options to given target items, and therefore, allows the (simulated) users to change their mind during the CRS interaction. We used crowdsourcing to  extend the existing well-used Shoes and FashionIQ with alternatives for 200 target items each. By using these extended datasets for evaluation, we show that previous (single-target) evaluations may underestimate the effectiveness of CRS systems on these datasets. Indeed, if they accept other alternative items, and are willing to switch strategy, then the system may satisfy them sooner. %
We also showed that introducing alternatives to the user simulator can lead to different conclusions, in the relative ranking of different CRS systems by effectiveness, and that users indeed tend to prefer alternatives when they have the option. As future work, we plan to use our alternatives for better evaluation of performance prediction techniques for image-based CRS.%

\bibliographystyle{splncs04}

\end{document}